\documentclass[aps,pre,twocolumn,showcaps,superscriptaddress,showpacs,amsmath,amssymb]{revtex4}
\usepackage{color}
\usepackage{graphicx}
\usepackage{dcolumn}
\usepackage{bm}
\usepackage[dvips]{epsfig}

\newcommand{\unit}[1]{\,\mathrm{#1}}
\newcommand{\mean}[1]{\langle #1 \rangle}

\newcommand{\kT}{\mathit{k_\mathrm{B}T}}
\newcommand{\ps}{p_\mathrm{s}}
\newcommand{\vloc}{v_\mathrm{s}}
\newcommand{\vvloc}{\vec{v}_\mathrm{s}}
\newcommand{\js}{j_\mathrm{s}}
\newcommand{\fc}{f_\mathrm{c}}
\newcommand{\vjs}{\vec{j}_\mathrm{s}}
\newcommand{\deff}{D_\mathrm{eff}}
\newcommand{\dnull}{D_0}
\newcommand{\peq}{p_\mathrm{eq}}
\renewcommand{\vec}[1]{\mathbf{#1}}

\begin{document}

\title{Characterizing Potentials by a Generalized Boltzmann Factor}

\author{V. Blickle}
\affiliation{2. Physikalisches Institut, Universit\"at Stuttgart,
Pfaffenwaldring 57, 70550 Stuttgart, Germany}

\author{T. Speck}
\affiliation{{II.} Institut f\"ur Theoretische Physik,
Universit\"at Stuttgart, Pfaffenwaldring 57, 70550 Stuttgart,
Germany}

\author{U. Seifert}
\affiliation{{II.} Institut f\"ur Theoretische Physik,
Universit\"at Stuttgart, Pfaffenwaldring 57, 70550 Stuttgart,
Germany}

\author{C. Bechinger}
\affiliation{2. Physikalisches Institut, Universit\"at Stuttgart,
Pfaffenwaldring 57, 70550 Stuttgart, Germany}

\begin{abstract}
  Based on the concept of a nonequilibrium steady state, we present a novel
  method to experimentally determine energy landscapes acting on colloidal
  systems. By measuring the stationary probability distribution and the
  current in the system, we explore potential landscapes with barriers up to
  several hundred $\kT$. As an illustration, we use this approach to measure
  the effective diffusion coefficient of a colloidal particle moving in a
  tilted potential.
\end{abstract}



\maketitle


{\it Introduction. --} The interaction of soft matter systems with potential
landscapes created by optical tweezers plays a key role for, e.g., mechanical
flexibility measurements of single biomolecules or molecular
motors~\cite{meh99,wan97}, guiding of neuronal cells~\cite{ehr02}, or phase
transitions of colloidal monolayers on patterned
substrates~\cite{man03,bru02}. In addition, extended optical lattices can be
used as sorters for microscopic particles~\cite{don03} or as
microoptomechanical devices such as Couette rheometers~\cite{lad05}.
Currently, no theories are available which can be used to directly calculate
optical trapping forces on macromolecules.  Thus the precise calibration of
optical forces is a central issue in many experiments.

The simplest method to determine an optical potential $V(\vec r)$ is to
measure the equilibrium distribution $\peq(\vec{r})$ of a highly diluted
colloidal system at position $\vec{r}$. From the inverted Boltzmann
factor
\begin{equation}
  V(\vec{r})=-\kT\ln\peq(\vec{r})
  \label{eq:boltzmann}
\end{equation}
one directly obtains the underlying potential $V(\vec r)$ with a typical
energy resolution on the order of $0.1\,\kT$~\cite{roh04,man03}, where
$k_\mathrm{B}$ is Boltzmann's constant and $T$ the temperature of the
surrounding fluid. This technique, however, is only applicable to potential
depths up to $\simeq7\,\kT$ which are effectively sampled by Brownian
particles in equilibrium. For larger trapping potentials, optical forces are
typically calibrated indirectly by taking advantage of Stokes law which
relates the particle velocity to the friction force exerted by the surrounding
solvent molecules. Accordingly, from the drift velocity of a particle, the
underlying potential can be reconstructed ~\cite{fauc95, pou97, ska06}.
Alternatively, within the drag force method, $V(\vec r)$ can be determined
from the particle's displacement upon moving the sample stage (and thus the
liquid) with known velocity~\cite{wan97,meh99,she98,ghi94}. In contrast to
Eq.~\eqref{eq:boltzmann}, however, the latter two nonequilibrium methods
neglect thermal fluctuations since only mean values of particle velocities or
displacements are considered. While such fluctuations can be neglected at
large trapping forces, this is no longer justified for external forces with
strengths comparable to those exerted by fluctuating Brownian forces.

In this paper, we introduce a potential reconstruction method based on a
generalization of Eq.~\eqref{eq:boltzmann} to nonequilibrium conditions.  This
is experimentally realized by generating a nonequilibrium steady state (NESS)
for a colloidal particle in a one-dimensional (toroidal) potential landscape.
By measuring the stationary probability distribution and the current in the
system, we can reliably calibrate potentials wells between a few tens up to
several hundreds of $\kT$.


{\it Potential reconstruction. --} Our method is based on a generalization of
the Boltzmann factor inversion~\eqref{eq:boltzmann} to nonequilibrium. The
effectively one-dimensional motion of the particle along a toroidal trap is
governed by a Langevin equation
\begin{equation}
  \dot{x}=\gamma^{-1}F(x)+ \zeta(t)
\end{equation}
with $x$ the spatial coordinate and $\zeta(t)$ representing thermal noise with
correlations $\mean{\zeta(t)\zeta(t')}=2(\kT/\gamma)\delta(t-t')$, where
$\gamma$ is the friction coefficient. The force $F(x)=-V'(x)+f$ exerted on the
particle stems from two sources, the gradient of the periodic potential
$V(x+L)=V(x)$ and a nonconservative driving force $f$.

We define a pseudo-''potential'' $\phi(x)$ by writing the nonequilibrium
steady state probability distribution as $\ps(x)=\exp[-\phi(x)]$ resembling
the Boltzmann factor. The stationary probability current through the toroid is
given as
\begin{equation}
  \label{eq:current}
  \js = \gamma^{-1}\left[F(x)\ps(x)-\kT\ps'(x)\right],
\end{equation}
which is constant in one dimension. We introduce the local mean velocity
\begin{equation*}
  \vloc(x) = \js/\ps(x)
\end{equation*}
and obtain~\cite{spec06}
\begin{equation}
  \label{eq:vloc}
  \gamma\vloc(x) = F(x) + \kT\phi'(x) = -V'(x) + f + \kT\phi'(x).
\end{equation}
Integration of Eq.~\eqref{eq:vloc} leads to the potential
\begin{equation}
  V(x) = \kT\phi(x) + \int_0^x[f-\gamma\vloc(x)]dx
  \label{Eq:Integral}
\end{equation}
up to an irrelevant additive constant. Using the definitions of $\phi(x)$ and
$\vloc(x)$ we finally arrive at
\begin{equation}
  V(x)=-\kT\ln\ps(x) + fx-\gamma\js\int_0^x\ps^{-1}(x)dx.
  \label{Eq:potential}
\end{equation}
Hence, the stationary probability $\ps(x)$ and the local mean velocity
$\vloc(x)$ determine the potential $V(x)$. The driving force $f$ can be
determined by setting $x=L$ in Eq.~\eqref{Eq:Integral} and using the
periodicity of the potentials $V(x)$ and $\phi(x)$ as
\begin{equation}
  f=\frac{\gamma}{L}\int_0^L\vloc(x) dx.
  \label{eq:force}
\end{equation}
In thermal equilibrium both $\js$ and $f$ vanish and Eq.~\eqref{Eq:potential}
reduces to the inverted Boltzmann factor Eq.~\eqref{eq:boltzmann}. Therefore
Eq.~\eqref{Eq:potential} can be understood as an extension of the Boltzmann
factor to nonequilibrium stationary states.


\begin{figure}[t]
  \includegraphics[width=\linewidth]{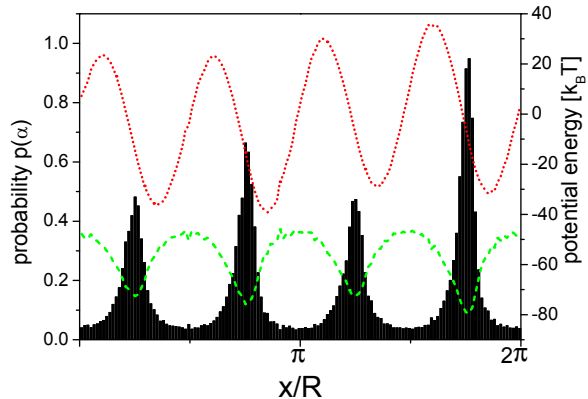}
  \caption{(color online) Stationary probability distribution $\ps(x)$ (black
    bars) and the pseudopotential $\phi(x)$ (dashed line), measured at a
    driving force of $34\unit{\kT /\mu m}$, pointing along the negative
    x-direction. The potential $V(x)$ (dotted line) is determined according to
    Eq.~\eqref{Eq:potential}. Note that $\phi(x)$is multiplied by a factor of
    10 and shifted vertically to enhance visualization.}
  \label{fig:potential}
\end{figure}

{\it Experiment. --} For an experiment exploiting Eq.~\eqref{Eq:potential}, we
use a scanning optical tweezers setup as described in detail
elsewhere~\cite{lutz06}. A laser beam ($\lambda\simeq532\unit{nm}$) is
deflected on a pair of galvanometric mirrors and focused with a 100\textbf{x},
NA=1.3 oil immersion objective from below onto a silica bead immersed in water
(diameter $d\simeq1.85\unit{\mu m}$). Upon periodic modulation of the angular
mirror positions we obtain a three dimensional toroidal laser trap with a
torus radius of $R\simeq3.95 \unit{\mu m}$.  At our driving frequencies
$\nu_{\mathrm T}\simeq 100 \unit{Hz}$, the particle can not follow directly
the rotating laser trap. Instead, every time the particle is passed by the
laser tweezers, it experiences a minute kick along the rotation direction
whose strength depends on the laser intensity $I_0$ \cite{fauc95}. Because the
particle´s trajectory is monitored with video microscopy at a sampling rate of
$20\unit{Hz}$, single kicking events are not resolved and the driving force
$f$ along the angular direction can be considered as constant~\cite{lutz06}.
In addition, the intensity of the laser is weakly modulated along the toroidal
trap. This is achieved with an electro-optical modulator (EOM) being
controlled by a function generator which is synchronized with the scanning
motion of the mirrors. This intensity modulation $I_m(x)$ leads to a periodic
potential $V(x)$ with $x$ the arc-length coordinate along the circumference of
the torus. It has been demonstrated that the resulting optical forces of such
an intensity modulated scanned laser tweezers exerted on a colloidal particle
correspond to those of a tilted periodic potential ~\cite{lutz06}.

\begin{figure}
  \includegraphics[width=0.9\linewidth]{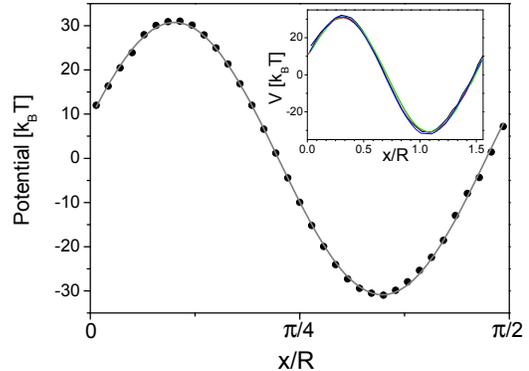}
  \caption{(color online) Averaged potential (solid points). The grey curve is
    a fit of Eq.~\eqref{Eq:V} to the averaged potential. Inset: Averaged
    potentials obtained for driving forces $f$ ($=34$, $43$, $57$ and
    $73\unit{\kT / \mu m}$).}
  \label{fig:pot}
\end{figure}

To experimentally demonstrate that $V(x)$ can be obtained under
non-equilibrium steady state conditions, the intensity of the scanned laser
tweezer along the toroidal trap was varied according to
\begin{equation}
  I(x) = I_0+I_{m}\sin(n\frac{2 \pi x}{L}). \hspace{0.7cm} (n=4)
  \label{eq:intensity}
\end{equation}
Fig.~\ref{fig:potential} shows the steady state probability distribution
$\ps(x)$ as obtained from the particle´s trajectory and the corresponding
pseudo-potential $\phi(x)$ for $I_0=44\unit{mW}$ and $I_{m}=10\unit{mW}$.
Together with the driving force $f=34\unit{\kT/\mu m}$ as determined from the
measured local mean velocity $\vloc(x)$ (cf. Eq.~\eqref{eq:force}) we finally
arrive at the potential $V(x)$ which is also plotted in
Fig.~\ref{fig:potential} as dotted line. Clearly, under NESS conditions the
minima and maxima of $\phi(x)$ and $V(x)$ do not coincide. In addition,
$\phi(x)$ varies in a less pronounced way than $V(x)$ because $\ps(x)$ is
broader than it would be in equilibrium. On top of the intensity modulation
according to Eq.~\eqref{eq:intensity} we observe a constant, small variation
of the potential with $2\pi$-periodicity caused by minute optical distortions
in our setup. Since we are only interested in the local shape of the
potential, in the following we only consider potentials, where $V(x)$ is
averaged over the four externally applied periods. The averaged potential,
which is plotted as solid bullets in Fig.~\ref{fig:pot}, is in excellent
agreement with a sinusoidal fit to
\begin{equation}
  V(x) = \frac{V_0}{2} \sin(4\frac{2 \pi x}{L})
  \label{Eq:V}
\end{equation}
as theoretically expected for the optical potential in case of sinusoidal
intensity variations~\cite{tlu98}.

To demonstrate the robustness of our approach in characterizing equilibrium
potentials under NESS conditions, we systematically varied the driving force
$f=34$, $43$, $57$ and $73\unit{\kT/\mu m}$ while keeping $V(x)$ unchanged.
Experimentally, this is achieved by changing $I_0$ with all other parameters
in Eq.~\eqref{eq:intensity} fixed. The measured potentials plotted in the
inset of Fig.~\ref{fig:pot} clearly fall on top of each other and thus
demonstrate that the measured $V(x)$ is independent of $f$.

\begin{figure}
  \includegraphics[width=0.9\linewidth]{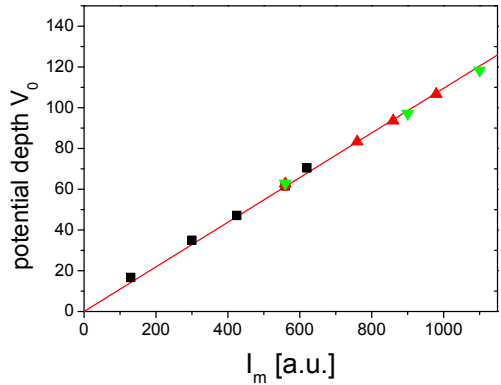}
  \caption{(color online) Potential depth as a function of the modulation
    amplitude $I_m$. The different symbols correspond to data acquired at
    different driving forces. ($\blacksquare$:$43$,
    \textcolor{red}{$\blacktriangle$}:$57$ and
    \textcolor{green}{$\blacktriangledown$}:$73\unit{\kT/\mu m}$) }
  \label{fig:ampli}
\end{figure}

Similarly, the potential amplitude $V_0$ can be changed by variation of $I_m$.
Fig.~\ref{fig:ampli} shows the potential depth as a function of $I_m$,
measured at different driving forces $f$ (marked by different symbols). As
expected, we find a linear dependence between $V_0$ and $I_m$ independent of
$f$. This shows that reconstruction of equilibrium potentials can be reliably
performed for a wide range of driving forces. For practical purposes, however,
the driving force should not exceed a certain range where the lower limit is
reached when the particle only rarely surmounts adjacent potential barriers
and thus cannot sample the entire landscape. For very large $f$, the
probability distribution becomes rather flat and very long sampling times are
required to accurately measure $V(x)$.


{\it Diffusion in tilted periodic potentials. --} Having demonstrated the
validity of our approach to reconstruct equilibrium potentials under NESS
conditions, in the following we will exemplarily apply this method to the
problem of giant diffusion. It has been shown theoretically~\cite{rei01}
and experimentally~\cite{lee06,tat03} that the effective diffusion coefficient
$\deff$ of a Brownian particle moving in a tilted periodic potential
$U(x)=V(x)+fx$ exhibits a pronounced maximum as a function of the driving
force $f$. Until now experiments were not able to match quantitatively the
theoretical predictions. With the ability to characterize the underlying
potential landscape in detail, we can quantitatively test the theoretical
behavior of $\deff$.

The effective diffusion coefficient is easily obtained from the particle
trajectory according to
\begin{equation}
  \deff = \lim_{t\rightarrow\infty}D(t), \qquad
  D(t) = \frac{\mean{x(t)^2}-\mean{x(t)}^2}{2t}.
  \label{Eq:Deff}
\end{equation}
This expression takes into account both the thermal diffusion and the drift
motion evoked by the tilt of the potential. It is therefore applicable to both
equilibrium and nonequilibrium conditions. Depending on the strength of the
driving force, three regimes can be distinguished: (i) At small $f$, the
particle is largely confined to the potential $V(x)$. Thus $\deff<\dnull$ with
$\dnull$ the diffusion coefficient of a free particle. (ii) Around a critical
force $\fc$, a considerable enhancement of the thermal diffusion occurs, i.e.
$\deff>\dnull$~\cite{rei01}. (iii) In the limit of very large $f$ the
potential becomes irrelevant and $\deff$ eventually approaches $\dnull$.

\begin{figure}[t]
  \includegraphics[width=\linewidth]{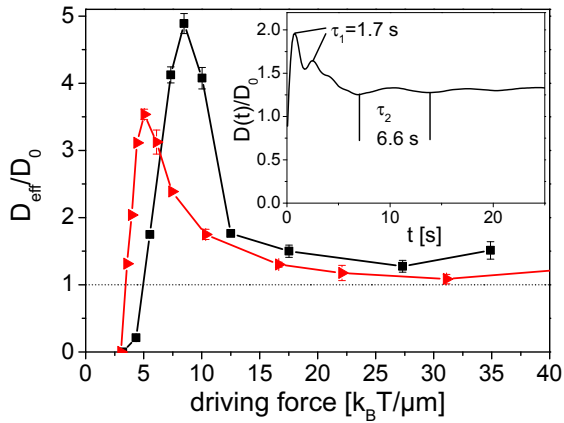}
  \caption{(color online) Normalized diffusion coefficient $\deff/\dnull$ vs
    external driving force. The data was obtained for potential depths $V_0$
    of $10.3$ (\textcolor{red}{$\blacktriangleright$}) and $14.4\unit{\kT}$
    ($\blacksquare$). Inset: The function $D(t)$ in Eq.~\eqref{Eq:Deff} versus
    time $t$ ($f\simeq6.6\unit{kT/\mu m}$, $V_0\simeq10.3\unit{\kT}$). After
    $15\unit{s}$ the long time limit is reached. The remaining small
    oscillations define the error in determining $\deff$.}
  \label{fig:giantdiff}
\end{figure}

Our results are shown in Fig.~\ref{fig:giantdiff}, where we have chosen the
same sinusoidal potential as above (see Eq.~\ref{Eq:V}) with typical
amplitudes between $10-20\,\kT$~\footnote{Deeper potentials are experimentally
  harder to explore because they lead to a rather sharp peak in $\deff(f)$.}.
Since the infinite time limit required to calculate $\deff$ cannot be realized
in experiments, we first plotted the right hand side of Eq.~\eqref{Eq:Deff} as
a function of time to determine when this expression saturates. The inset of
Fig.~\ref{fig:giantdiff} shows the result obtained for $f\simeq6.6\unit{kT/\mu
  m}$ and $V_0\simeq10.3\,\kT$. After an initial peak, the curve converges to
the corresponding long-time value. A closer inspection reveals two damped
oscillations whose periods are easily explained: The short oscillation time
$\tau_1\simeq1.7s$ corresponds to the mean residence time of the particle
within one minimum while the other oscillation with
$\tau_2\simeq6.6\unit{s}\simeq4\tau_1$ equals the mean revolution time of the
particle along the torus. After about $t\gtrsim15\unit{s}$, both oscillations
have essentially decayed to the long-time value corresponding to $\deff$.

\begin{table}[b]
  \begin{tabular}{c||c|c||c|c}
    $V_0$ [$\kT$] & \multicolumn{2}{c||}{$f_{c}[\unit{\kT/\mu m}]$}
                  & \multicolumn{2}{c}{$D/\dnull$}\\
                  & exp. & theo. & exp. & theo.\\
    \hline
    \hline
    14.4 & 8.5 & 7.3 & 4.9 & 3.1 \\
    10.3 & 5.1 & 5.2 & 3.5 & 2.5 \\
  \end{tabular}
  \caption{\label{tab:tab}
    Comparison between experimentally determined and
    theoretically predicted position and height of the giant diffusion
    peak.}
\end{table}

Fig.~\ref{fig:giantdiff} shows the normalized effective diffusion coefficients
for potential depths of $10.3$ and $14.4\,\kT$. Both curves show a peak
clearly indicating the enhancement of thermal diffusion in tilted periodic
potentials. With increasing potential strength we observe a shift of the
$\deff(f)$ curve towards larger forces. The values of $\fc$ and $\deff(\fc)$
sensitively depend on the shape of the potential and are theoretically
predicted as $\fc=2V_0/R$ and $\deff(\fc)=0.0696
\dnull(\frac{2}{3}V_0\pi^3)^{(2/3)}$~\cite{rei01}.

A comparison with our data is shown in Tab.~\ref{tab:tab}. While the predicted
critical force $\fc$ is in rather good agreement with the experimental data,
the theoretical values systematically underestimate $\deff(\fc)/\dnull$ by a
factor of about $0.7$. The origin of this discrepancy is due to the
aforementioned slight distortions along the toroidal trap which leads to local
variations in the potential depth and thus affects the effective diffusion
coefficient. Since at the same time the local shape of $V(x)$ is hardly
affected by those distortions, the good agreement for the critical tilt can be
explained.


{\it Concluding perspective. --} So far, we have demonstrated a novel method
to reconstruct equilibrium potentials on the basis of the stationary
probability distribution. In particular in one-dimensional NESS conditions,
this quantity is easily determined experimentally, because the stationary
current $\js$ is constant. When the method is extended to higher dimensions in
the presence of nonconservative force fields $\vec f(\vec r)$, in addition to
the steady state probability $\ps(\vec r)$ the local mean velocity
$\vvloc(\vec r)=\vjs(\vec r)/\ps(\vec r)$ is required. Experimentally, this
quantity is obtained by averaging the velocity of particles passing $\vec r$.
Then the actual potential could be reconstructed through integration along
open paths $C(\vec r)$ starting at an arbitrary but fixed initial point and
ending in $\vec r$, leading to
\begin{multline}
  V(\vec r) = -\kT\ln\ps(\vec r) \\
  + \int_{C(\vec r)}\left[\vec f(\vec r(s))
    -\gamma\vvloc(\vec r(s))\right]\cdot d\vec r(s).
\end{multline}

In summary, we have demonstrated a flexible method to characterize potentials
using the generalization of the inverted Boltzmann factor. In contrast to
equilibrium measurements, this allows to characterize laser potentials up to
depths of several hundredth or even thousandths of $\kT$. Based on the
determination of the stationary state probability distribution $\ps$, this
technique is easily applicable to different situations, e.g. topographical
potentials and does not require fast data acquisition techniques.


\end{document}